# Internet of Things Technologies for Managing COVID-19 Pandemic: Recommendations and Proposed Framework

Navod Neranjan Thilakrathne [1,*], Dr. Rohan Samarasinghe [2]

[1,2]Department of ICT, Faculty of Technology, University of Colombo

navod.neranjan@ict.cmb.ac.lk*

**Abstract:** The Internet of Things, often known as IoT, is an innovative technology that connects digital devices all around us, allowing Machine to Machine (M2M) communication between digital devices all over the world. Due to the convenience, connectivity, and affordability, this IoT is being served in various domains including healthcare where it brings exceptional benefits to improve patient care, uplifting medical resources to the next level. Some of these examples include surveillance networks, healthcare delivery technologies, and smart thermal detection. As of now, the IoT is served in various aspects of healthcare making many of the medical processes much easier as opposed to the earlier times. One of the most important aspects that this IoT can be used is, managing various aspects of healthcare during global pandemics, as pandemics can bring an immense strain on healthcare resources, during the pandemic. As there is no proper study is done with regards to the proper use of IoT for managing pandemics, in this regard, through our study we aim to review various use cases of IoT towards managing pandemics especially in terms of COVID-19; owing to what we are currently going through. In this regard, we are proposing a conceptual framework synthesizing the current literature and resources, which can be adopted when managing global pandemics to accelerate the battle pace with these deadly pandemics and focusing on what the entire world is currently going through where almost more than four (04) million people are diminished of this COVID-19 pandemic.

**Background**

The Internet of Things (IoT), connects all of the things around us, creating a massive omnipresent network that allows these linked things to communicate with one another [1]. As of now, this IoT is highly used in healthcare owing to the enormous benefits it yields, which act as the main enabler for global healthcare [2]-[5]. Through the immense benefits that IoT offers, it is evident that it gives us the flexibility for managing the lack of medical resources such as infrastructure and medical staff, management of chronic diseases, solving medical issues associated with aging populations, and managing pandemic events [6]. During the course of a pandemic, the smart services offered by IoT can be used towards providing ***public health surveillance, supply chain management, discovering the source (patient zero) of the pandemic, community support, tracking patients, monitoring the patient condition, providing treatments for patients and capturing and analyzing real-time data***; which is significant for mitigating and controlling most of the adverse effects associated with pandemics, that can control the pandemic to an acceptable or a minimum level, which eventually result in eradicating the pandemic entirely from this world, with the establishment of proper world-wide public health surveillance systems with the collaboration of relevant stakeholders [7]- [15]. Following we highlight some of the key application use cases that IoT can be used during the pandemic, with the aim of reducing the strain on medical resources, for the better understanding of our readers.

- *Remote patient monitoring*
  IoT powered remote monitoring or well known as Telemedicine helps medical staff and caregivers to monitor patient conditions remotely, where most of the IoT devices first gather pathological details from patients and send them to medical staff and caregivers in remote locations for their recommendation, attention, and treatment, so they can take timely actions [1]-[3]. This is highly useful in monitoring the condition of patients, who are located in rural areas during a pandemic; that are otherwise hard to reach or monitor the condition of contagious patients maintaining appropriate social distancing.

- *Medical management*
  In terms of the management of healthcare resources during a pandemic, the scalability of IoT is playing a vital role when monitoring patients remotely and managing medical resources such as hospital beds. Gauteng health services in South Africa, for example, implemented an Electronic Bed Management System (eBMS) to determine the availability of beds across several facilities in order to determine resource availability [4],[5]-[10]. Further as maintaining social distancing is vital during a virus outbreak daily patient check-ups can be easily done using this IoT, as patients themselves can take their temperature, blood pressure and upload the data to the cloud for the attention of caregivers even when they are at their homes. This will also reduce the workload of medical staff and reduce the possibility of cross-infection of medical staff during a pandemic.

- *Ensure compliance with quarantine*
  IoT could be used to ensure patient compliance with quarantine rules once the potentially infected people have been quarantined. Officials from the Department of Public Health will keep track of which patients have been quarantined and which have not. The IoT data would also help them figure out who else might have been affected [1]-[4], and keep track of patients whereabouts in real time, who are under quarantine [10].

- *Medical robots*
  IoT-powered medical robots are often used in clinical settings to perform tasks that are otherwise hard to perform for medical staff such as performing high-precision surgeries. This enables us to achieve medical procedures with high accuracy. It is evident that medical robots are currently being deployed in COVID-19 detection centers to automate most of the operations and replace humans to avoid health workers getting cross infected [5]-[10].

Having provided a preamble about IoT, next COVID-19 stands for "Coronavirus Disease-2019," a respiratory ailment caused by the severe acute respiratory syndrome coronavirus-2 (SARS-CoV-2). SARS-CoV-2, like the typical influenza virus, targets the respiratory system and causes symptoms such as cough, fever, tiredness, and shortness of breath. This was originally reported in December 2019 in Wuhan City, Hubei Province, China. Since then, the COVID-19 has spread like wildfire over the rest of the world in a much shorter time, resulting in the death of more than 04 million people. COVID-19, like all other virus epidemics, poses significant hurdles, including identifying the source of the pandemic (patient zero), minimizing virus dissemination, and incorporating sufficient medical resources to treat all patients with severe symptoms in a timely manner [15]-[20]. On the other hand, the demand for medical resources would also be increased, owing to the strain created by the pandemic. Hence the management of pandemics clearly necessitates the identification of infected persons, as well as the tracking and implementation of relevant measures such as social distancing and the use of Personal Protective Equipment (PPE) [20]-[26].

Recent COVID-19 experiences demonstrate the need of taking a wise and speedy approach to dealing with pandemics to prevent putting an undue burden on healthcare resources and reducing loss of human lives [27]. Since the discovery of COVID-19, it has been able to spread across all continents, causing many challenges highlighting the need for scalable ways to mitigate such kinds of pandemics. It is clear that tracing the source of an outbreak, quarantining potentially infected patients, treating seriously ill patients, and preventing cross-infection between medical staff and patients all require a significant amount of human resources, highlighting the necessity for a practical and scalable solution, where this IoT comes in to play owing to its ubiquitous nature [27]-[32].

**Objectives**

The main objective of our research is to review the prominent IoT applications used to curb the COVID-19 pandemic; identify the main challenges and issues arising with the incorporation of IoT technologies to tackle the COVID-19 pandemic and proposes a generic conceptual framework that provides a generic integration of IoT technologies which can be adopted by relevant stakeholders such as policy-making bodies like World Health Organisation (WHO), individual nations and their responsible bodies (e.g.: ministry of health, Srilanka) as per their discretion; which can be used to manage the COVID-19 pandemic to an acceptable level which eventually results in eradicating the pandemic entirely from the world. Not only that, but we also believe this could be adopted to manage other pandemics as well, with a little amount of customization where only minor changes needed to be done as COVID-19 is a highly contagious disease as opposed to other types of a pandemic; for an instance, Ebola which spread by direct contact with blood or other body fluids.

**Methodology**

This includes a keyword search for our review, where two sets of keywords were selected to search for available latest articles, where the IoT technology adaptation took place to curb the spread of the pathogen. The keywords were derived considering the latest trends and the hypes; as IoT is being a trending technology and the name COVID-19 or coronavirus we could hear daily, as we all are still being suffered from it. ***The first set of keywords were (Coronavirus or COVID-19 and IoT). The second set of keywords were (Coronavirus or COVID-19 and IoT and adoption or adaptation).*** Then we executed this to identify journal articles, conference papers/proceedings and, books paired with a selected news article from reputable sources, and a set of other published documents by health authorities globally in ***Scopus, Web of Science, and Google Scholar databases.***

**Proposed conceptual framework**

From the review we have done, we were able to constitute the main building blocks of the proposed framework [27]-[40]. The framework aims to develop an outline of three phases that can serve as a reference for future research and adopted by relevant stakeholders as we emphasized on the objectives section; to use IoT technologies for fighting COVID-19 as depicted in Fig.1. Our proposed framework composed of three phases

***Phase one:*** This phase represents the input of the framework which comprises COVID-19 contaminants factors with IoT technologies. In addition, ***stakeholders (WHO, government, medical organizations and etc.) responsibilities*** are also presented which act as the input for our framework.

***Phase two:*** This includes major IoT applications which could be used to contain the COVID-19 outbreak, which we discussed some applications already. These applications could be carried out by incorporating various IoT technologies. The implementation of these technologies may initiate various issues and challenges, which are also represented in the second phase.

***Phase three or the output***: Successful implementation of these technologies by the relevant stakeholders can promote an improvement in the efficacy of healthcare providers by a reduction in the imposed workload. Nevertheless, it also will result in reduced mistakes, improved diagnosis, improved performance, effective management, enhanced treatment, and finally lesser cost when dealing with the pandemic control; where it will result in controlling the pandemic to an acceptable level, and eventually result in eradicating the pandemic. On the other hand, the relevant stakeholders can define their own evaluation criteria and key performance indicators towards evaluating their implementations (e.g.: how many quarantine patients were advised through telemedicine on a one particular day).

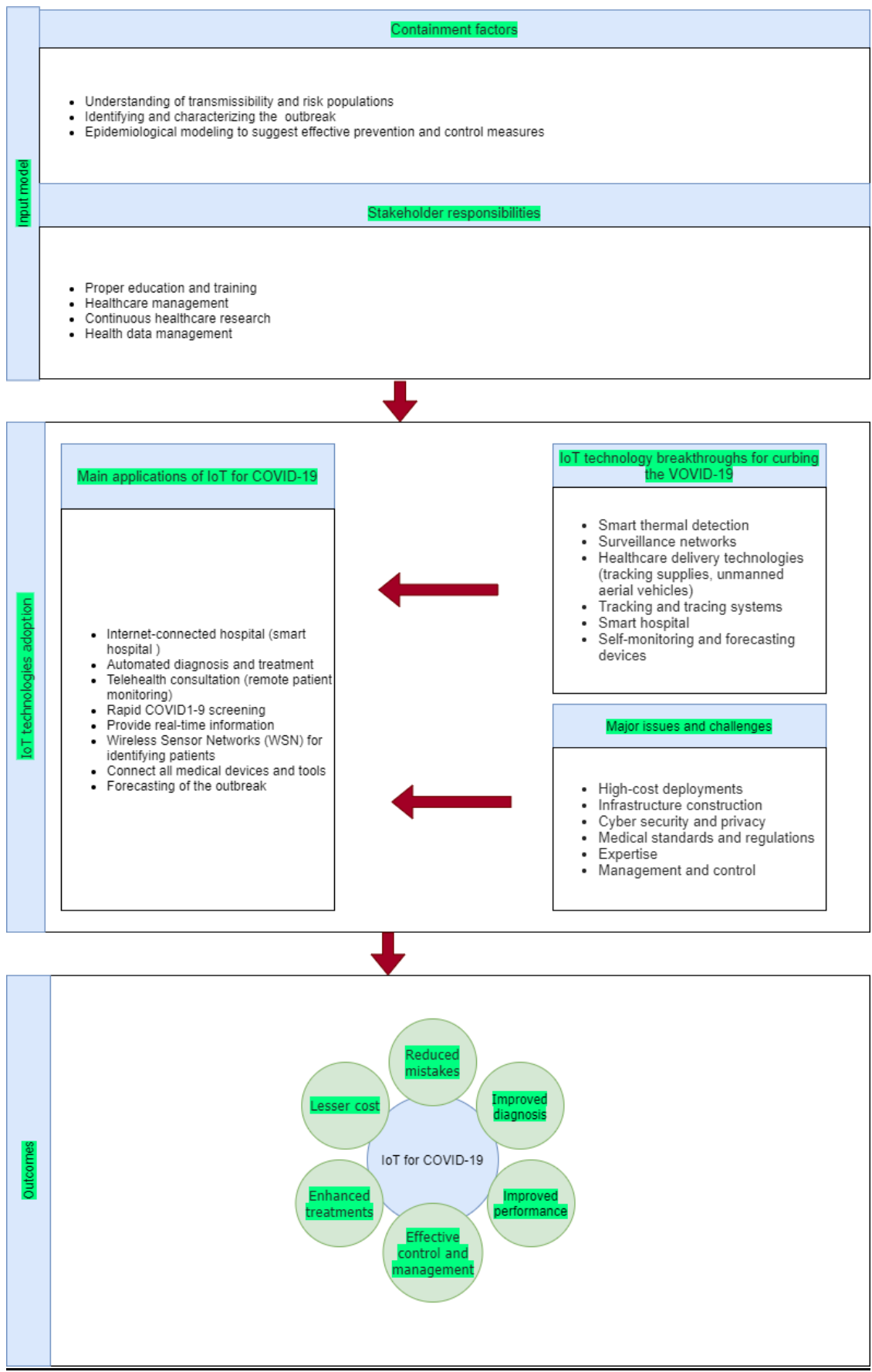

**Fig. 1** Proposed framework

**Recommendations**

a) For policy makers and planners

As the main objective of our research is to review the prominent IoT applications used to curb the COVID-19 pandemic; identify the main challenges and issues arising with the incorporation of IoT technologies to tackle the COVID-19 pandemic and proposes a generic conceptual framework that provides a generic integration of IoT technologies which can be adopted by relevant stakeholders such as policy-making bodies like World Health Organisation (WHO), individual nations and their responsible bodies (e.g.: ministry of health, Srilanka) as per their discretion; which can be used to manage the COVID-19 pandemic to an acceptable level which eventually results in eradicating the pandemic entirely from the world. Not only that, but we also believe this could be adopted to manage other pandemics as well, with a little amount of customization where only minor changes needed to be done as COVID-19 is a highly contagious disease as opposed to other types of a pandemic; for an instance, Ebola which spread by direct contact with blood or other body fluids.

b) For researchers

As the occurrence of the next pandemic is imminent and unpredictable, we can never be certain of when or where the next pandemic will arise, where early preparedness is essentially required. Hence we believe as per future research, future research should more focus on strengthening public health surveillance, where it would help to contain the virus outbreak with less effort, in a short time without allowing the outbreak to grow into a mega pandemic.

**Conclusion**

Based on the review we have done, it is evident that the adoption of IoT in underdeveloped and developing countries towards pandemic management is really less as opposed to developed countries, where it can attributed to not having enough capital for smart IoT deployments. Also based on the facts that we gathered, we suggest that while there have been implementations of IoT technologies in most of the countries; early detection and tracing, and tracking, the incorporation of more IoT-based technologies in more sectors is suspected to yield better results. Furthermore, Reports attempting to curb the spread of the COVID-19 infections using IoT technologies are on the rise, with reports of promising results. Therefore, we recommend adopting these IoT smart technologies, as much as possible to manage the pandemic to an acceptable level, where the contributions from individual stakeholders on whole would result in eradicating the virus outbreak entirely from world. Lastly, we need to know that occurrence of the next pandemic is imminent and unpredictable. Hence, we can never be certain of when or where the next pandemic will arise, where early preparedness is essentially needed. We hope this research will be useful for healthcare professionals, researchers, academics, students, and anyone who seeks new knowledge in the area of ICT in healthcare.